\documentclass{article}
\usepackage{amsmath, amssymb, amsfonts}
\title{On a de Sitter-like spacetime with cylindrical symmetry} 
\author{Hristu Culetu\\ Ovidius University, Dept.of Physics and Electronics, \\B-dul Mamaia 124, 900527 Constanta, Romania \\ email : hculetu@yahoo.com}
\begin{document}
\numberwithin{equation}{section}
\pagenumbering{arabic}
\maketitle
\begin{abstract}
A curved static de Sitter-like metric is analyzed. The source of curvature is rooted from a constant stress tensor with positive energy density and negative pressures. All the curvature invariants are constant everywhere and the geometry is conformally flat. The horizon surface gravity equals the parameter $\omega$ from the metric that is also interpreted as an angular velocity. The Tolman-Komar gravitational energy is investigated. One finds that the horizon entropy satisfies the relation $S = A_{H}/4$, as for the black hole horizon.
\end{abstract}

\newcommand{\fv}{\boldsymbol{f}}
\newcommand{\tv}{\boldsymbol{t}}
\newcommand{\gv}{\boldsymbol{g}}
\newcommand{\OV}{\boldsymbol{O}}
\newcommand{\wv}{\boldsymbol{w}}
\newcommand{\WV}{\boldsymbol{W}}
\newcommand{\NV}{\boldsymbol{N}}
\newcommand{\hv}{\boldsymbol{h}}
\newcommand{\yv}{\boldsymbol{y}}
\newcommand{\RE}{\textrm{Re}}
\newcommand{\IM}{\textrm{Im}}
\newcommand{\rot}{\textrm{rot}}
\newcommand{\dv}{\boldsymbol{d}}
\newcommand{\grad}{\textrm{grad}}
\newcommand{\Tr}{\textrm{Tr}}
\newcommand{\ua}{\uparrow}
\newcommand{\da}{\downarrow}
\newcommand{\ct}{\textrm{const}}
\newcommand{\xv}{\boldsymbol{x}}
\newcommand{\mv}{\boldsymbol{m}}
\newcommand{\rv}{\boldsymbol{r}}
\newcommand{\kv}{\boldsymbol{k}}
\newcommand{\VE}{\boldsymbol{V}}
\newcommand{\sv}{\boldsymbol{s}}
\newcommand{\RV}{\boldsymbol{R}}
\newcommand{\pv}{\boldsymbol{p}}
\newcommand{\PV}{\boldsymbol{P}}
\newcommand{\EV}{\boldsymbol{E}}
\newcommand{\DV}{\boldsymbol{D}}
\newcommand{\BV}{\boldsymbol{B}}
\newcommand{\HV}{\boldsymbol{H}}
\newcommand{\MV}{\boldsymbol{M}}
\newcommand{\be}{\begin{equation}}
\newcommand{\ee}{\end{equation}}
\newcommand{\ba}{\begin{eqnarray}}
\newcommand{\ea}{\end{eqnarray}}
\newcommand{\bq}{\begin{eqnarray*}}
\newcommand{\eq}{\end{eqnarray*}}
\newcommand{\pa}{\partial}
\newcommand{\f}{\frac}
\newcommand{\FV}{\boldsymbol{F}}
\newcommand{\ve}{\boldsymbol{v}}
\newcommand{\AV}{\boldsymbol{A}}
\newcommand{\jv}{\boldsymbol{j}}
\newcommand{\LV}{\boldsymbol{L}}
\newcommand{\SV}{\boldsymbol{S}}
\newcommand{\av}{\boldsymbol{a}}
\newcommand{\qv}{\boldsymbol{q}}
\newcommand{\QV}{\boldsymbol{Q}}
\newcommand{\ev}{\boldsymbol{e}}
\newcommand{\uv}{\boldsymbol{u}}
\newcommand{\KV}{\boldsymbol{K}}
\newcommand{\ro}{\boldsymbol{\rho}}
\newcommand{\si}{\boldsymbol{\sigma}}
\newcommand{\thv}{\boldsymbol{\theta}}
\newcommand{\bv}{\boldsymbol{b}}
\newcommand{\JV}{\boldsymbol{J}}
\newcommand{\nv}{\boldsymbol{n}}
\newcommand{\lv}{\boldsymbol{l}}
\newcommand{\om}{\boldsymbol{\omega}}
\newcommand{\Om}{\boldsymbol{\Omega}}
\newcommand{\Piv}{\boldsymbol{\Pi}}
\newcommand{\UV}{\boldsymbol{U}}
\newcommand{\iv}{\boldsymbol{i}}
\newcommand{\nuv}{\boldsymbol{\nu}}
\newcommand{\muv}{\boldsymbol{\mu}}
\newcommand{\lm}{\boldsymbol{\lambda}}
\newcommand{\Lm}{\boldsymbol{\Lambda}}
\newcommand{\opsi}{\overline{\psi}}
\renewcommand{\tan}{\textrm{tg}}
\renewcommand{\cot}{\textrm{ctg}}
\renewcommand{\sinh}{\textrm{sh}}
\renewcommand{\cosh}{\textrm{ch}}
\renewcommand{\tanh}{\textrm{th}}
\renewcommand{\coth}{\textrm{cth}}

\section{Introduction}
The observed accelerating expansion (at an increasing rate) of our universe is consistent with a positive cosmological constant $\Lambda \approx 10^{-123}$ in Planck units. There is considerable evidence that our universe is asymptotically de Sitter \cite{TP1, AS} ($\Lambda > 0$). According to Padmanabhan, the expansion of the universe (viewed as an emergence of space) is driven towards holographic equipartition when it becomes de Sitter and the holographic principle may be written as $N_{sur} = N_{bulk}$ \cite{TP1} ($N_{sur}$ is the number of degrees of freedom (DOF) on the Hubble surface of radius $H^{-1}$ and $N_{bulk} = 2|\bar{E}|/k_{B}T$ is the number of DOF which are in equipartition at the horizon temperature $T = H/2\pi$ if $|\bar{E}|$ is taken to be the Tolman-Komar (TK) energy \cite{TP2}).

Because of the negative pressures, the TK energy of the de Sitter universe is negative (we remind that the energy density and pressures contribute in the same manner in the expression of the gravitational energy). Thanks to the vanishing of the timelike Killing vector at $r = H^{-1}$, the de Sitter horizon is endowed with a temperature given by $T = H/2\pi$ and an entropy $S = A_{H}/4 = \pi H^{-2}$, so that $|\bar{E}| = 2TS = H^{-1}$.

We investigate in this work a curved space with cylindrical symmetry to which the $z = const$ surface is equivalent with the equatorial plane $\theta = \pi/2$ for the de Sitter space in its static coordinates. We found that the source of curvature is an anisotropic fluid with positive constant energy density and constant negative pressures. As for de Sitter geometry, the TK energy is negative and the relation $|\bar{E}| = 2TS$ is satisfied. Even though our spacetime has no spherical symmetry, the entropy is given by one quarter from the horizon area and the equation $N_{sur} = N_{bulk}$ is obeyed. 

The units are chosen such that $c = G = k_{B} = \hbar = 1$.  

\section{Anisotropic stress tensor}
We propose the de Sitter-like metric with axial symmetry to be given by
  \begin{equation}
ds^{2} =-(1 - \omega^{2}r^{2}) dt^{2} + \frac{dr^{2}}{1 - \omega^{2}r^{2}} + dz^{2} + r^{2} d\phi^{2}
\label{2.1}
\end{equation}
where $(t, r, z, \phi)$ are the usual cylindrical coordinates, $\omega$ is a positive constant and $r < 1/\omega$. To be an exact solution of Einstein's equation $G_{ab} = 8\pi T_{ab}$, the following (diagonal) stress tensor is to be put on its r.h.s.
  \begin{equation}
  T^{t}_{~t} = -\rho = \frac{-\omega^{2}}{8 \pi},~~~ T^{r}_{~r} = p_{r} = \frac{-\omega^{2}}{8 \pi},~~~T^{z}_{~z} = p_{z} = \frac{-3\omega^{2}}{8 \pi},~~~ T^{\phi}_{~\phi} = p_{\phi} = \frac{-\omega^{2}}{8 \pi},
\label{2.2}
\end{equation}
(all the other components are vanishing). We observe that $T^{a}_{~b}~ (a,b = 0, 1, 2, 3)$ is not of $\Lambda$ - form (no Lorentz invariance) due to the $p_{z}$ pressure which is not equal with $p_{r}$ or $p_{\phi}$. Strictly speaking, 
$T^{a}_{~b}$ cannot be Lorentz invariant (in spite of its constancy) because of the preffered location of the z - axis. Not only $T^{a}_{~b}$ is finite everywhere but also the curvature invariants $R^{a}_{~a} = 6\omega^{2}$, $R^{ab}R_{ab} = 12\omega^{4}$ and the Kretschmann scalar $R^{abcd}R_{abcd} = 12\omega^{4}$. It is worth noting that the Weyl tensor $C_{abcd} = 0$, so the metric (2.1) is conformally-flat. This is not trivial even though the de Sitter space is also conformally-flat, in its time-dependent version. The usual coordinate transformation \cite{RT}
  \begin{equation}
  \bar{r} = \frac{r}{\sqrt{1 - \omega^{2}r^{2}}} e^{-\omega t},~~~\bar{t} = t + \frac{1}{2\omega} ln(1 - \omega^{2}r^{2})
\label{2.3}
\end{equation}
cannot bring the metric (2.1) in the well-known time dependent (comoving) de Sitter form (and, further, in its conformally flat form) because of the axial symmetry of the former (we see that $\omega$ plays here the role of the Hubble constant). 

As for the de Sitter geometry the metric (2.1) is endowed with an event horizon at $r_{H} = 1/\omega$. The horizon represents a surface of a cylinder with the radius $r_{H}$ and length, say,  $\Delta z = z_{2} - z_{1}$. Its area is $2\pi r_{H} \Delta z = (2\pi/\omega) \Delta z$, if we take $z \in (z_{1}, z_{2})$. The area is, of course, infinite if $z$ span the entire $z$ - axis.

\section{Congruence of static observers}
Let us consider now a congruence of static observers given by the velocity vector field
  \begin{equation}
  u^{b} = \left(\frac{1}{\sqrt{1 - \omega^{2}r^{2}}}, 0, 0, 0\right).
\label{3.1}
\end{equation}
The acceleration 4-vector appears as 
  \begin{equation}
   a^{b} \equiv  u^{a} \nabla_{a} u^{b} = (0, -\omega^{2}r, 0, 0)  
\label{3.2}
\end{equation}
with $\sqrt{a^{b} a_{b}} = \omega^{2}r/\sqrt{1 - \omega^{2}r^{2}}$. Since $a^{r} = -\omega^{2}r$ is negative, the field is repulsive (to maintain ourselves at a static position one has to accelerate towards the z-axis). Using (3.2) we could compute now the surface gravity $\kappa$ on the horizon. We have
  \begin{equation}
  \kappa = \sqrt{a^{b} a_{b}}~ \sqrt{-g_{tt}}|_{H} = \omega,
\label{3.3}
\end{equation}
so that one possible meaning of $\omega$ may rely on the horizon surface gravity. In that case, for example, $\rho$ becomes proportional to an acceleration squared, a natural interpretation in Newtonian gravity.

We might also associate an Unruh temperature $T = \omega/2\pi$ to the horizon $r_{H} = 1/\omega$. It can be written as, including fundamental constants, $k_{B}T = \hbar \omega/2\pi$, whence $\omega$ may be interpreted as a mean frequency and $k_{B}T$ as a mean energy per particle from the Unruh radiation. 

We could also consider $\omega$ to be an angular velocity. We cojecture therefore that the spacetime (2.1) corresponds to an uniformly rotating observer, with the constant frequency $\omega$ (for example, an observer fixed on some rigid disk with radius $r < 1/\omega$). In other words, a curved geometry is generated by rotation, the source of the field being given by the stress tensor (2.2). Hence, the metric viewed by a uniformly rotating observer cannot be obtained from the Minkowski metric by a coordinate transformation. Padmanabhan \cite{TP3} has noticed that gauge degrees of freedom (DOF) are upgraded to true (physical) DOF by infinitesimal coordinate transformations. We assume a similar effect takes place in our situation: physical DOF are created by rotation so that we find in a bath of constant classical energy density $\rho$ spread everywhere, like a cosmological constant. We mention also an analogy with the Unruh effect where particles are generated out of vacuum due to the accelerated motion.

The fact that the parameter $\omega$ may be considered as a frequency of rotation can be justified as follows. From (2.1) we have
  \begin{equation}
-1 =-(1 - \omega^{2}r^{2}) \dot{t}^{2} + \frac{\dot{r}^{2}}{1 - \omega^{2}r^{2}} + \dot{z}^{2} + r^{2} \dot{\phi}^{2}
\label{3.4}
\end{equation}
where $\dot{t} = dt/d\tau$, etc., $\tau$ being the proper time. Take now a test particle at $r = r_{0} = const., z = const.$. Eq. (3.4) yields now
  \begin{equation}
-1 =-(1 - \omega^{2}r_{0}^{2}) \dot{t}^{2} + r_{0}^{2} \dot{\phi}^{2}
\label{3.5}
\end{equation}
Cylindrical symmetry and the static character of the metric (2.1) gives us
  \begin{equation}
  \dot{t} = \frac{E}{1 - \omega^{2}r_{0}^{2}},~~~~\dot{\phi} = \frac{L}{r_{0}^{2}},
\label{3.6}
\end{equation}
where $E > 0$ and $L$ are, respectively, the energy per unit mass and the angular momentum per unit mass of the test particle. From (3.5) and (3.6) one obtains
  \begin{equation}
-1 =-\frac{E^{2}}{1 - \omega^{2}r_{0}^{2}} + \frac{L^{2}}{r_{0}^{2}} 
\label{3.7}
\end{equation}
which yields
  \begin{equation}
\frac{E^{2} - 1 + \omega^{2}r_{0}^{2}}{1 - \omega^{2}r_{0}^{2}} = \frac{L^{2}}{r_{0}^{2}}. 
\label{3.8}
\end{equation}
It is known that $E = 1$ for the black hole geometry corresponds to a particle dropped from infinity with zero velocity into the black hole \cite{SS}. In our case ''near infinity'' means ''far from the horizon''. We put therefore $E = 1$ in (3.8) to obtain 
  \begin{equation}
\frac{ \omega^{2}r_{0}^{2}}{1 - \omega^{2}r_{0}^{2}} = \frac{L^{2}}{r_{0}^{2}} 
\label{3.9}
\end{equation}
Taking now $\omega r_{0} << 1$, we get $L = \omega r_{0}^{2}$ which shows $\omega$ is the angular velocity of the test particle describing a circular trajectory of radius $r_{0}$.

\section{The Tolman-Komar energy}
Let us find now the Tolman-Komar gravitational energy \cite{TP2, HC} for the spacetime (2.1)
   \begin{equation}
W = 2 \int(T_{ab} - \frac{1}{2} g_{ab}T)u^{a} u^{b} N\sqrt{\gamma} d^{3}x ,
\label{4.1}
\end{equation}
where $N = \sqrt{-g_{tt}} = \sqrt{1 - \omega^{2}r^{2}}$ is the lapse function and $\gamma$ is the determinant of the spatial 3-metric. By means of $u^{a}$ from (3.1) and $T^{a}_{~b}$ from (2.2), one finally obtains
\begin{equation}
W = - \frac{\omega^{2}r^{2}}{2} \Delta z,
\label{4.2}
\end{equation}
where the integral over $z$ is taken for $z\in (z_{1}, z_{2}), ~\phi \in (0, 2\pi)$ and $W$ gives the energy enclosed by a cylinder of length $\Delta z$ and radius $r$. In terms of $p_{z}$ from (2.2), the above expression can be written as $W = (4\pi/3) r^{2} p_{z} \Delta z = - (4\pi/3) r^{2} \sigma$, where $p_{z} \Delta z = - \sigma$ could be taken as a surface pressure ($\sigma$ - the surface tension on a tape of radius $r$ and width $\Delta z$). $W$ is negative due to the negative pressures giving their contribution to the total gravitational energy. This is obvious for a de Sitter-like spacetime. The result (4.2) is similar with Padmanabhan's one, given by $E = -H^{2}R^{3}$ \cite{TP2}, valid for spherical symmetry (his $H$ is Hubble's constant and $R$ - the sphere radius). The differences are rooted from the axial symmetry used in our case. $W$ is finite everywhere and on the horizon acquires the very simple form $W = -(1/2)\Delta z$. When the fundamental constants are introduced, we obtain $W_{H} = -(c^{4}/2G) \Delta z$. The constant force $F = -c^{4}/2G$ appears here as an attractive force due to the horizon itelf. In other words, one seems that $a^{r} = -\omega^{2}r$ is negative because the test particle points towards the horizon. A similar expression for the above attractive force has been found out by Easson et al. \cite{EFS}. Their entropic force $F_{r}$ points in the direction of increasing  entropy or the screen, which in their case is the horizon. It is worth to stress that $F$ is a radial force because we integrated $W$ w.r.t. $r\in (0, r_{H})$, $\Delta z$ being constant.
It is instructive noting that the energy $W_{H}$ contained in the whole cylinder of radius $r_{H}$ and height $\Delta z$ seems to originate from the mass $M = (c^{2}/2G)\Delta z$ which leads to $\Delta z = 2GM/c^{2}$. Namely, $\Delta z$ is equivalent to the gravitational radius of the mass $M$. 

Using the previous expression of the horizon temperature and the thermodynamic relation $|W| = 2TS$ \cite{TP2} we may evaluate now the horizon entropy $S$
\begin{equation}
S = \frac{|W|}{2T} = \left( \frac{\omega^{2}r^{2}}{2} \Delta z \frac{\pi}{\omega} \right)|_{H} = \frac{\pi}{2\omega} \Delta z.
\label{4.3}
\end{equation}
Keeping in mind that the horizon area is $A_{H} = 2\pi r_{H} \Delta z = (2\pi/\omega) \Delta z$, one obtains $S = A_{H}/4$, as for the black hole horizon. Hence, the above result is preserved for cylindrical symmetry.

Keeping in mind that the de Sitter universe is related to expansion and our de Sitter-like metric (2.1) to rotation, we may assume that we have here an equivalence between expansion and rotation.

Let us estimate the main physical parameters for $\omega = 1 rad/s$. The energy density becomes $\rho = \omega^{2}c^{2}/8\pi G \approx 10^{27} erg/cm^{3}$, the horizon radius $R_{H}$ is located at $3.10^{10} cm$ (the ''light cylinder'' is a true horizon in our situation). For the surface gravity we have $\kappa = |a^{r}|_{H} = 3.10^{10} cm/s^{2}$ and the horizon temperature is $T = \hbar \omega/2\pi k_{B} \approx 10^{-12} K$. Being extensive quantities, $W$ and $S$ depend also on the values spanned by the coordinate $z$, i.e. by $\Delta z$.

\section{Conclusions}
There are experimental evidences that the expansion of our universe is accelerating and asymptotically de Sitter. Therefore, we have good reasons to investigate further the de Sitter space to find new properties. The $r - t$ part of our metric (2.1) is of de Sitter type but, nevertheless, it possesses cylindrical symmetry. Its curvature is generated by an anisotropic energy-momentum tensor $T^{a}_{~b}$ with constant energy density and pressures. $T^{a}_{~b}$ is not of the $\Lambda$ - form because of the symmetry axis and, therefore, it is not Lorentz-invariant. As the de Sitter space, our spacetime (2.1) is endowed with an event horizon at $r_{H} = 1/\omega$ where the parameter $\omega$ is considered to be a frequency of rotation. 

We found also that the horizon entropy $S$ obeys the equation $S = A_{H}/4$, where $A_{H}$ is the horizon area enclosed by two $z = const.$ planes. The TK energy is negative and proportional to $r^{2}$ but the total energy depends only on $\Delta z$, the interval spanned by the coordinate $z$.


\begin{thebibliography}{8}


\bibitem{TP1}
T. Padmanabhan, arXiv: 1206.4916 [gr-qc].
\bibitem{AS}
A. Sheykhi, arXiv: 1304.3054 [gr-qc].
\bibitem{TP2}
T. Padmanabhan, Class. Quant. Grav. 21, 4485 (2004) (arXiv: gr-qc/0308070).
\bibitem{RT}
R. C. Tolman, Relativity, Thermodynamics and Cosmology, Oxford University Press, 1934, pp. 347.
\bibitem {HC}
H. Culetu, Int. J. Mod. Phys.: Conf. Series, 3, 455 (2011) (arXiv: 1101.2980 [gr-qc]); arXiv: 1303.7376 [gr-qc].
\bibitem{TP3}
T. Padmanabhan, arXiv: 1302.1206 [gr-qc].
\bibitem{SS}
M. Smerlak and S. Singh, arXiv: 1304.2858 [gr-qc].
\bibitem{EFS}
D. Easson, P. Frampton and G. Smoot, Phys. Lett. B696, 273 (2011) (arXiv: 1002.4278 [hep-th].

\end{thebibliography}
\end{document}